\documentclass[conference]{IEEEtran}
\IEEEoverridecommandlockouts
\usepackage{cite}
\usepackage{amsmath,amssymb,amsfonts}
\usepackage{algorithmic}
\usepackage{graphicx}
\usepackage{textcomp}
\usepackage{xcolor}
\usepackage{multirow}
\usepackage{hyperref}
\usepackage{booktabs}
\usepackage{tabularx}
\usepackage{caption}
\usepackage{subcaption}

\def\BibTeX{{\rm B\kern-.05em{\sc i\kern-.025em b}\kern-.08em
    T\kern-.1667em\lower.7ex\hbox{E}\kern-.125emX}}
\begin{document}

\title{Can CAV Reduce Non-Recurrent Urban Road Congestion?\\

}

\author{\IEEEauthorblockN{Yunkai Li, Haotian Li}
\IEEEauthorblockA{\textit{Beijing-Dublin International College} \\
\textit{University College Dublin}\\
Dublin, Ireland \\
\{yunkai.li, haotian.li\}@ucdconnect.ie}
\and
\IEEEauthorblockN{Beatriz Martinez-Pastor}
\IEEEauthorblockA{\textit{School of Civil Engineering} \\
\textit{University College Dublin}\\
Dublin, Ireland \\
beatriz.martinez-pastor@ucd.ie}
\and
\IEEEauthorblockN{Shen Wang}
\IEEEauthorblockA{\textit{School of Computer Science} \\
\textit{University College Dublin}\\
Dublin, Ireland \\
shen.wang@ucd.ie}
}

\maketitle

\begin{abstract}
A well-designed resilient and sustainable urban transportation system can recover quickly from the non-recurrent road congestion (NRC), which is often caused by en-route events (e.g., road closure due to car collisions). Existing solutions, such as on-board navigation systems and temporary rerouting road signs, are not effective due to delayed responses. Connected Autonomous Vehicles (CAV) can be helpful in improving recurrent traffic as they can autonomously adjust their speed according to their real-time surrounding traffic, sensed by vehicular communications. Preliminary simulation results in this short paper show that CAV can also improve traffic when non-recurrent congestion occurs. Other results in fuel consumption, CO\textsubscript{2} emission, and traditional traffic safety indicators are open for future discussions.
\end{abstract}


\section{Introduction}
Unlike recurrent urban road congestion that is mainly caused by the high traffic demand at specific hours of the day, NRC \cite{skabardonis2003measuring} is difficult to predict in advance and requires timely actions before it gets propagated in greater areas. A resilient \cite{nogal2016resilience} and sustainable urban transportation system should recover quickly from the NRC, which is often caused by en-route events (e.g., road closure due to car collisions). Existing solutions such as onboard navigation systems \cite{wang2016next} and temporary rerouting road signs, are not effective due to excessively delayed responses. The emerging Connected Autonomous Vehicles (CAV) are likely to be widely used on the road soon for improving traffic efficiency and safety. CAV can sense their surrounding traffic (e.g., distance and relative speed to the other surrounding vehicles) in real-time, as they are equipped with vehicular communication devices using IEEE 802.11p or 5G millimeter wave. Based on the sensed traffic, CAV can also drive autonomously in adjusting their best speeds to keep a safe distance while achieving stable and relatively high speed. Existing research \cite{gueriau2020quantifying} revealed the great potential of CAV to improve the recurrent traffic in many ways. However, the influence of CAV on the non-recurrent one still remains unclear. This paper, to the best of our knowledge, is the first to study the impact of CAV on releasing the NRC under various penetration rates. The preliminary simulation results show that under a small, large grid, and a realistic urban scenario of Dublin city centre, CAV demonstrate their superiority in reducing travel time, fuel consumption, CO\textsubscript{2} emission and improving traffic safety, with the growing penetration rate. Therefore, the resilience of the traffic system is increased when NRC occurs.

\section{Simulation Settings}
We use SUMO\footnote{https://www.eclipse.org/sumo/} to perform the simulation study. We mainly compare four cases: original (ORG) case with 100\% human-driven vehicles (HDV), ORG case with 100\% CAV, non-recurrent congestion (NRC) case with 100\% HDV, and NRC case with 100\% CAV. These four cases are assessed under each of three testing scenarios: small grid (3x4), large grid (20x20), and Dublin city centre (a subset of an open data\footnote{https://github.com/maxime-gueriau/ITSC2020\_CAV\_impact}) to test the CAV impact under various scalability. The traffic on each scenario is set to have congestion, but not heavily congested to leave room for improvement after NRC. The traffic generation for each scenario lasts 3600 seconds, and from 1200th second to the 2400th second, we close the road in the centre of gird maps, and four bridges in the Dublin city centre to mimic the NRC. We use the car following model Krauss to simulate HDV and IDM to simulate Level 4 CAV as specified in \cite{gueriau2020quantifying}. The vehicular communication is assumed to be perfect at the moment and the rerouting is set as the shortest-distance routing to enable quick response to NRC. We use the mean travel speed for each road, mean travel time for each vehicle, total fuel consumption and CO\textsubscript{2} emission for all vehicles to evaluate the traffic efficiency. To assess the traffic safety, we count the number of times when time-to-collision (TTC) is over 1.5s for HDV and 0.75s for CAV, this threshold setting keeps the same as \cite{gueriau2020quantifying}.
\begin{table}[h]
\centering
\caption{Main results. (The results of last three cases are shown in percentage changes compared to ORG)}
\begin{tabularx}{0.5\textwidth}{p{0.4cm}|p{1.15cm}|X|X|X|p{0.98cm}}
\toprule
                            &          & Mean TravelTime & Fuel Consumption & CO\textsubscript{2} Emission & TTC \\
\midrule
\multirow{3}{*}{\rotatebox[origin=c]{90}{\parbox[c]{1cm}{\centering Small Grid}}} 
                            & ORG      &       127.74s       &      573.84l    &   1334.94kg   & 8138  \\
                            & NRC      &       154.45\%      &      146.88\%   &   146.87\%    & 23.51\%  \\
                            & ORG(CAV) &       -9.82\%       &      -21.88\%   &   -21.88\%   & -48.89\%  \\
                            & NRC(CAV) &       -7.75\%       &      -20.31\%   &   -20.31\%   & -41.77\%  \\
\midrule
\multirow{3}{*}{\rotatebox[origin=c]{90}{\parbox[c]{1cm}{\centering Large Grid}}} 
                            & ORG      &       556.93s       &      5087.02l    & 11834.20kg   & 19147  \\
                            & NRC      &       32.97\%       &      -0.08\%     & -0.08\%      & -1.34\%  \\
                            & ORG(CAV) &       -5.38\%       &      -28.46\%    &   -28.46\%   & -38.44\%  \\
                            & NRC(CAV) &       -5.34\%       &      -28.48\%    &   -28.48\%   & -38.84\%  \\
\midrule
\multirow{3}{*}{\rotatebox[origin=c]{90}{\parbox[c]{1cm}{\centering Dublin Centre}}} 
                            & ORG      &       583.08s       &      6390.38l    & 14865.48kg   &  21413   \\
                            & NRC      &       3.15\%        &      2.14\%    & 2.14\%   &  6.36\%   \\
                            & ORG(CAV) &       -54.75\%      &      -64.75\%    & -64.75\%    &  27.44\% \\
                            & NRC(CAV) &       -54.43\%      &      -67.91\%    & -67.91\%    &  36.16\%  \\
\bottomrule
\end{tabularx}
\label{tab:sum_res}
\end{table}

\section{Evaluation Results}
\begin{figure}[h]
\begin{subfigure}{.24\textwidth}
  \centering
  % include first image
  \includegraphics[width=0.8\linewidth]{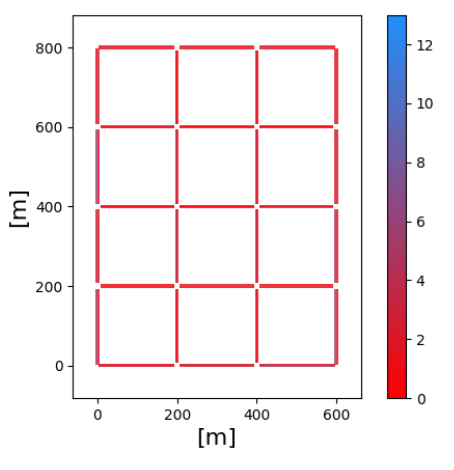} 
  \includegraphics[width=0.85\linewidth]{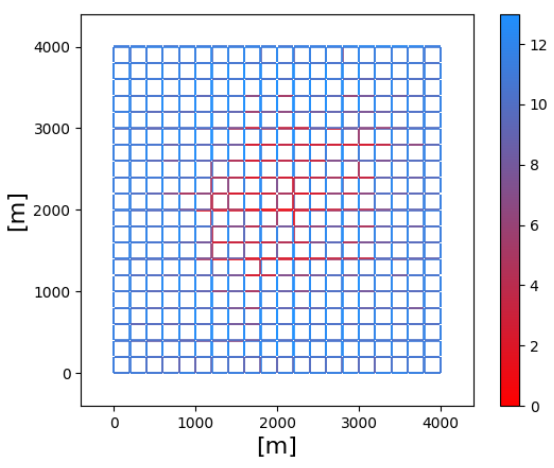}
  \includegraphics[width=1.\linewidth]{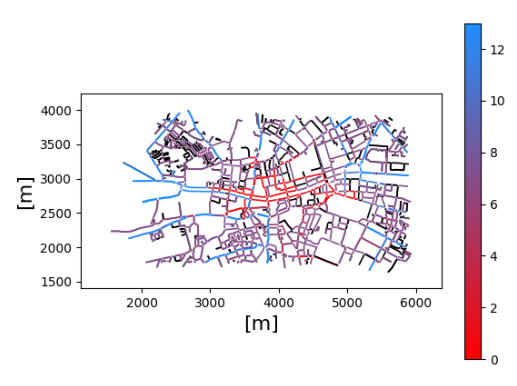}
  \caption{NRC(100\%HDV)}
  \label{fig:sub-first}
\end{subfigure}
\begin{subfigure}{.24\textwidth}
  \centering
  % include second image
  \includegraphics[width=0.8\linewidth]{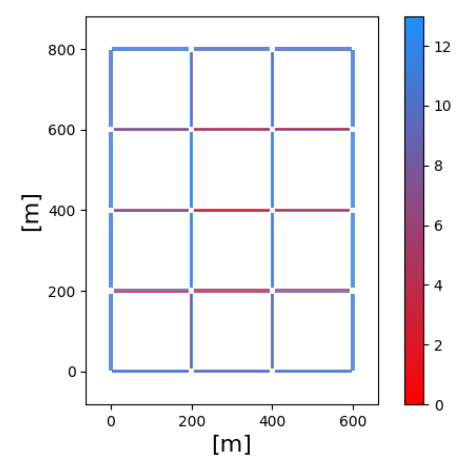}
  \includegraphics[width=0.85\linewidth]{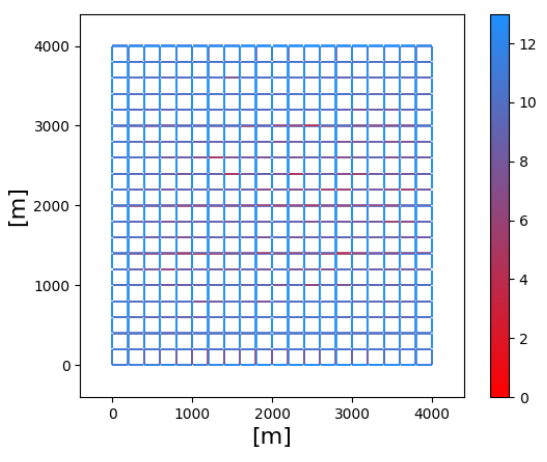}
  \includegraphics[width=1.\linewidth]{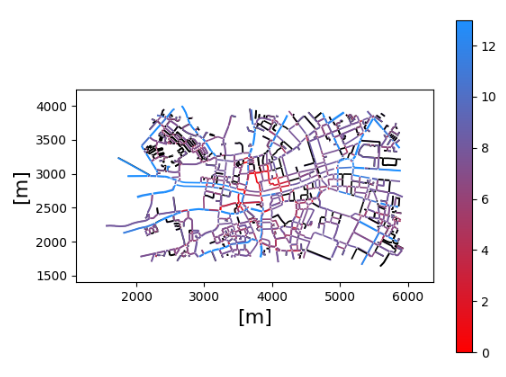} 
  \caption{NRC(100\%CAV)}
  \label{fig:sub-second}
\end{subfigure}
\caption{Heatmap of the average travel speed (m/s) over the whole simulation duration on each road of grid maps and Dublin centre.}
\label{fig:heat_map}
\end{figure}

We can see clearly from Figure \ref{fig:heat_map} that CAV greatly increase the average travel speed for roads around the closed central areas. In smaller areas such as the small grid scenario, CAV can almost improve the travel speed for the whole network. Although the closure of few central roads only deteriorates the traffic in their vicinity, as shown in the large grid and Dublin centre scenario, CAV can still largely narrow the NRC impacted area in terms of travel speed.

Moreover, as illustrated in Table \ref{tab:sum_res}, we can see that even when compared with the ORG where no roads are closed, NRC(CAV) can significantly reduce the travel time, fuel consumption, CO\textsubscript{2} emission, and improve road safety. The only exception is for TTC results in the Dublin centre scenario. All three cases compared to ORG, including all CAV cases, have increased by a fair amount. We think this is because CAV has better control of speed and distance, so they can drive at a higher speed while still keep a smaller distance with other vehicles. This happens more often when the road capacity is decreased due to the occurrence of NRC at complicated urban road networks where roads are short and curvy. We have also seen a similar result when assessing using another common traffic safety indicator called post-encroachment time (PET). Thus, a more fair traffic indicator might be required in the future for CAV in particular. Another interesting result is that for the large grid and Dublin centre scenario, road closure with HDV or CAV can even reduce the fuel consumption and CO\textsubscript{2} emission slightly. This finding needs more exploration but for the time being it demonstrates that reducing travel time even a significant amount, does not even guarantee a reduction in fuel consumption and CO\textsubscript{2} emission.

\begin{figure}[h]
    \centering
    \includegraphics[scale=0.23]{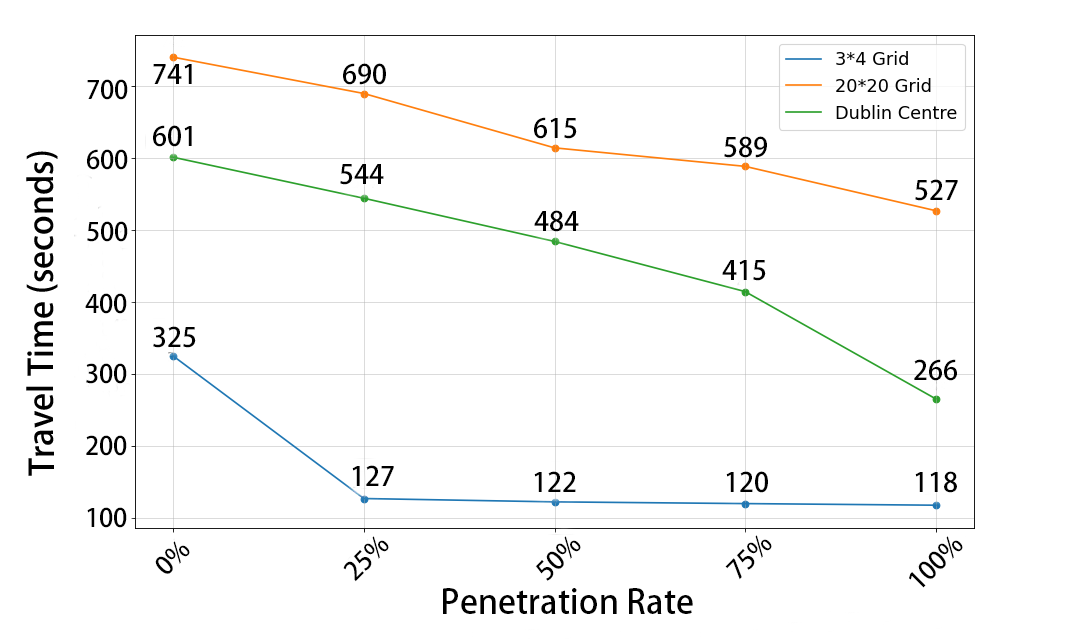}
    \caption{The average travel time under various penetration rate of CAV in three testing scenarios.}
    \label{fig:cav_ratio}
\end{figure}

Lastly, as shown in Figure \ref{fig:cav_ratio}, the main traffic indicator, travel time is decreased gradually with the growing CAV penetration rate. This encourages us to promote the introduction of CAV, even a small portion increase (e.g., 25\% in the small grid) can also clearly benefit the traffic.

\section{Conclusion and Future work}
The preliminary simulation results show that the more CAV are introduced on the urban road, the better the NRC can be alleviated in terms of travel time. Future work includes a better traffic safety metric for CAV and a smarter CAV rerouting to further alleviate NRC including emissions and fuel consumption. 

\bibliographystyle{IEEEtran}
\bibliography{references.bib}

\end{document}